\documentclass[twocolumn,showpacs,preprintnumbers,amsmath,amssymb,footinbib]{revtex4}

\usepackage{graphicx}
\usepackage{dcolumn}
\usepackage{bm}
\usepackage{url}

\begin{document}

\preprint{manuscript}

\title{Dense packing crystal structures of physical tetrahedra}

\author{Yoav Kallus}
\author{Veit Elser}
\affiliation{%
Laboratory of Atomic and Solid-State Physics, Cornell University, Ithaca, New York, 14853
}

\date{\today}

\begin{abstract}
We present a method for discovering dense packings of general convex
hard particles and apply it to study the dense packing behavior of
a one-parameter family of particles with tetrahedral symmetry
representing a deformation of the ideal mathematical tetrahedron
into a less ideal, physical, tetrahedron and all the way to the sphere. 
Thus, we also connect the two well studied
problems of sphere packing and tetrahedron packing on a single axis.
Our numerical results uncover a rich optimal-packing behavior,
compared to that of other continuous families of particles previously
studied. We present four structures as candidates for the optimal packing
at different values of the parameter, providing an atlas of
crystal structures which might be observed  in systems of
nano-particles with tetrahedral symmetry.
\end{abstract}

\maketitle

Impenetrable (hard) mathematical bodies
(e.g. spheres, spheroids, superballs,
and polyhedra) have received much attention as models for the
equilibrium behavior of systems of nano-particles and for the wealth of
equilibrium and non-equilibrium structures they exhibit \cite{spheroid,superball,Bezdek,GlotzerSolomon,torqnature,torqPRE}.
For the tetrahedron alone, quasicrystal structures,
novel crystal structures, and glassy structures have been reported
in numerical simulations \cite{torqnature,torqPRE,kallusdcg,Glotzer,chendcg}.
However, in all cases, the tetrahedral particles studied were
mathematically ideal (polyhedral)
tetrahedra. By contrast, in an experiment which found
that regular tetrahedra have random packings that are denser
than known for any other body, the tetrahedral macro-particles (dice)
used had rounded edges and vertices \cite{chaikinPRL}. Tetrahedral
nano-particles used as colloids are not only imperfectly-shaped
tetrahedra, but are also soft, in the sense that the interactions
beyond hard-core repulsion are significant \cite{nanotet,nanotet2}.
In this paper, we attempt to characterize the packing behavior of
physical, rather than mathematical,
tetrahedra by studying a one-parameter
family of particles with tetrahedral symmetry that interpolates
between the mathematical tetrahedron on one end
of the parameter's range
and the sphere on the other end. We explore the effect of this
parameter by constructing, using a {\it de novo}
numerical search, candidate structures
for the optimal packing of the different particles in the family.

\begin{figure}
\begin{center}
\includegraphics[trim=105 380 160 100,clip,scale=0.08]{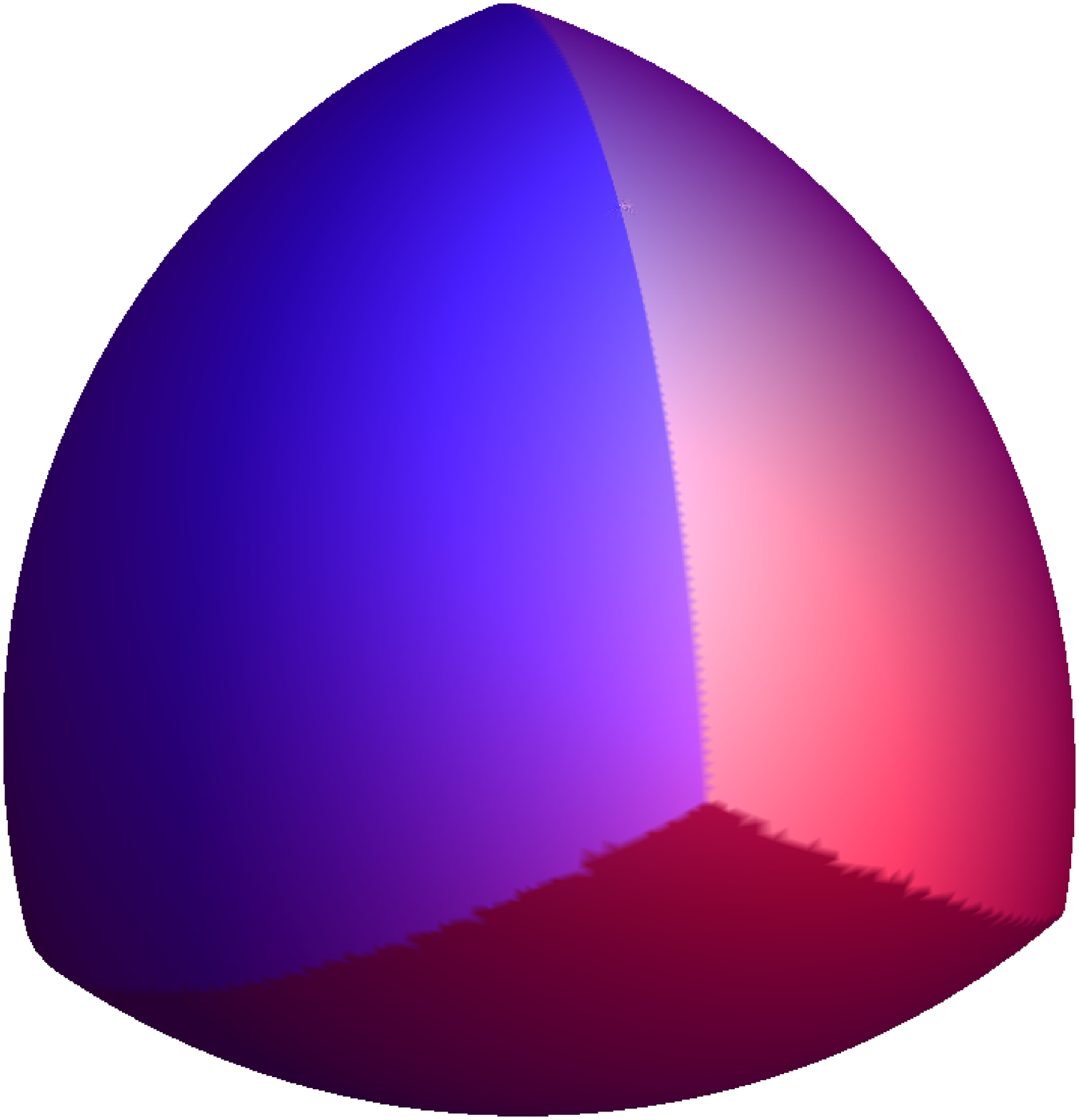}
\includegraphics[trim=105 380 160 100,clip,scale=0.08]{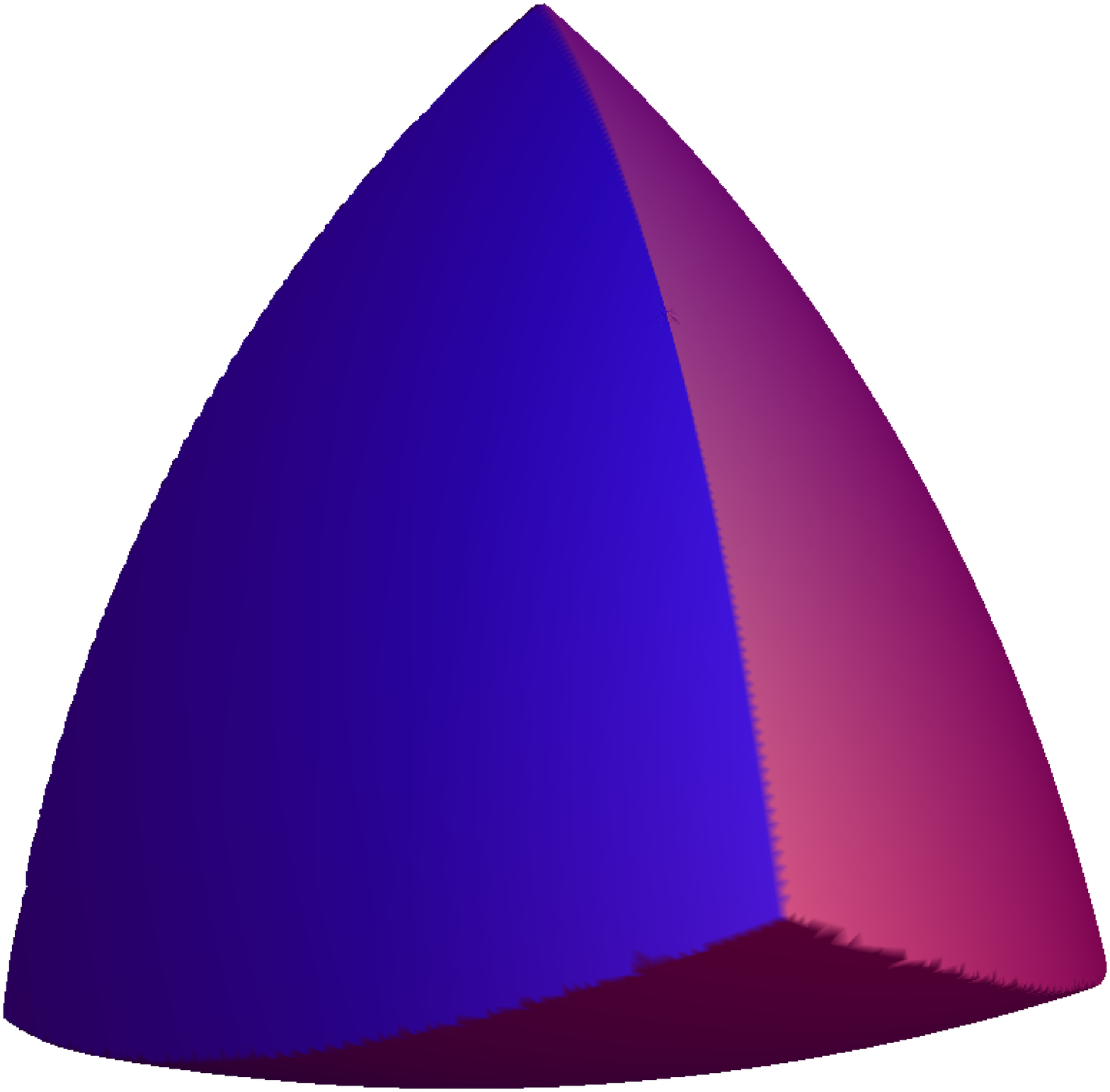}
\includegraphics[trim=105 380 160 100,clip,scale=0.08]{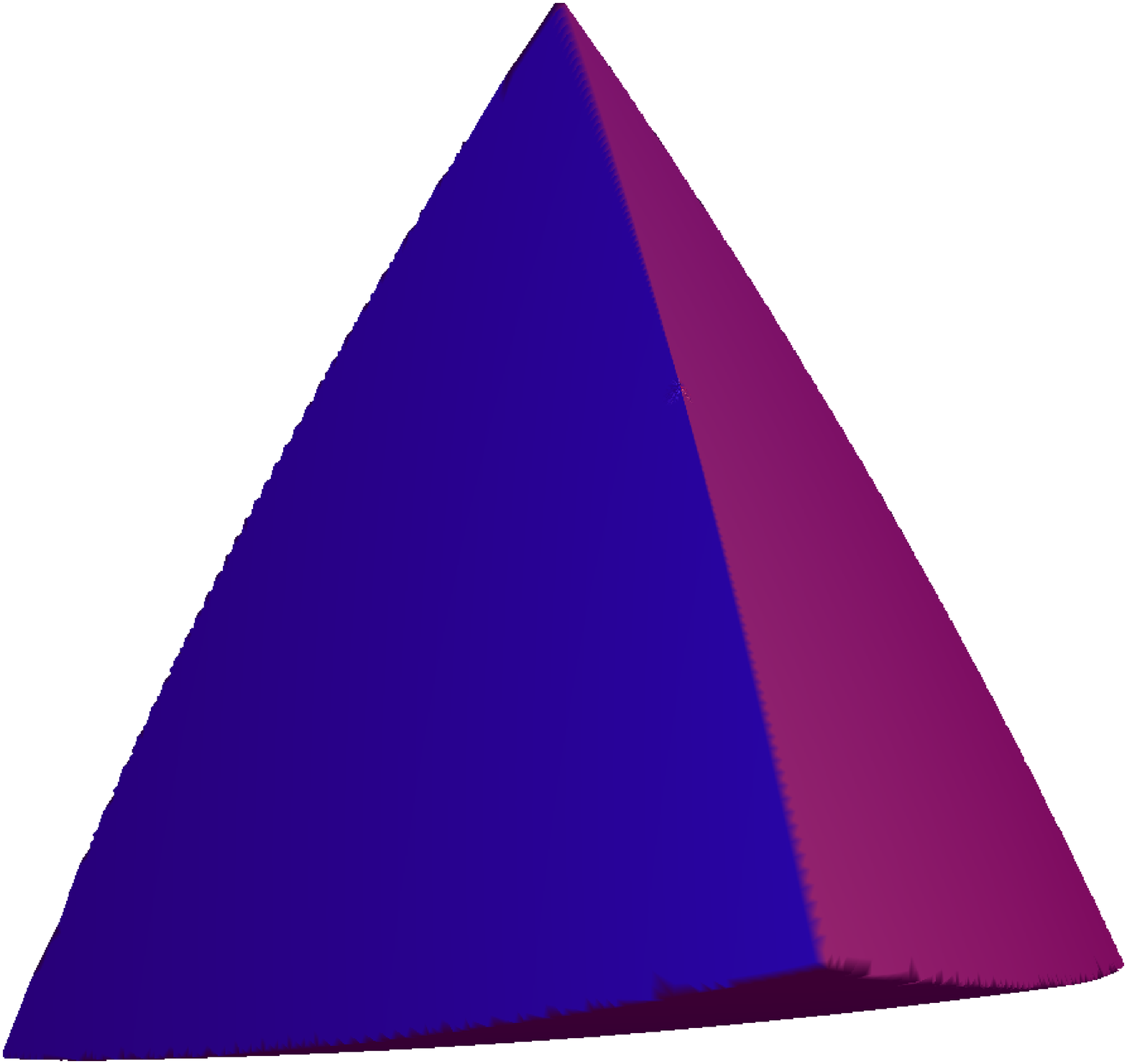}
\end{center}
\caption{\label{puffs} (Color online)
tetrahedral puffs of varying asphericity. From left to right, the asphericities
of the puffs shown are $\gamma=4/3,2,8/3$.}
\end{figure}

A particle interpolating between the sphere and the regular tetrahedron
can be achieved by a variety of constructions. The simplest construction,
probably, is to place the centers of four unit
spheres at the vertices of a regular tetrahedron with edges of length $a$
and consider the volume at the intersection of all four spheres.
We call the resulting figure a {\it tetrahedral puff} (Figure \ref{puffs}).
For a special value of $a$, the tetrahedral puff is the Reuleaux
tetrahedron (a three-dimensional version of the Reuleaux triangle,
but not a solid of constant width) \cite{Meissner}.
A more convenient parameter than this edge length is the asphericity
$\gamma$, which is the ratio between the radii of the
particle's circumscribing and inscribed spheres.
The value $\gamma=1$ obtained when the four
spheres coincide corresponds to
a sphere. The value $\gamma=3$, which is the largest asphericity possible for
a convex particle with tetrahedral symmetry and corresponds to a regular
tetrahedron, is obtained in the limit that the four spheres intersect at a point.
The Reuleaux tetrahedron is the puff with asphericity 
$\gamma=(3+\sqrt{24})/5\approx1.58$.

To efficiently search for candidate optimal packing
structures we run a numerical search at increasing
densities. To prevent overlaps between different
particles in the structure we employ a recently developed
overlap resolution technique. In a packing of congruent
particles, every particle can be obtained from a single
primitive particle $K$ by a rotation and a translation:
$K_1=\mathrm{R}_1 K + \mathbf{r}_1 = \{\mathrm{R}_1
\mathbf{x}+\mathbf{r}_1|\mathbf{x}\in K \}$,
where $\mathrm{R}_1$, a rotation matrix, and
$\mathbf{r}_1$, a translation vector, parameterize the
configuration of the particle $K_1$. We call an
{\it exclusion projection} the operation of, for any two
particles $K_1$ and $K_2$ that overlap, identifying the
new set of configuration parameters
$(\mathrm{R}'_1,\mathrm{R}'_2,\mathbf{r}'_1,\mathbf{r}'_2)$ that
resolves the overlap while minimizing the distance
in configuration-space from the original configuration,
$(\mathrm{R}_1,\mathrm{R}_2,\mathbf{r}_1, \mathbf{r}_2)$.
This is the projection in configuration-space to the set
of non-overlapping configurations. Below we give the method of
implementing this projection.

At high densities, however, we may have to resolve overlaps of a
particle with multiple other particles, and the projection that applies
to pairs cannot be used directly. Therefore, we introduce multiple
independent copies (replicas) of the
configuration parameters of each particle, so that after an exclusion
projection, any pair of particles will have at least one overlap-free
pair of replicas. Another projection ensures that in the final structure
obtained from the search all replicas of a single particle agree on its
configuration. This scheme, called {\it divide and concur} ($D-C$),
has previously allowed us to study dense packings of a variety of polyhedral
and high-dimensional spherical particles, and is here generalized to any
convex particles. Apart from the exclusion projection, described 
here, the rest of the application of the scheme to dense
periodic packing discovery is described in Ref. \cite{PDC}.

Given a pair of particles (or replicas) in a configuration
$(\mathrm{R}_1,\mathrm{R}_2,\mathbf{r}_1,\mathbf{r}_2)$
which overlaps, the exclusion projection 
resolves the overlap by identifying
a new set of configuration parameters 
$(\mathrm{R}'_1,\mathrm{R}'_2,\mathbf{r}'_1,\mathbf{r}'_2)$
while minimizing the distance to the original configuration as defined by
\begin{align}
d^2=&\sum_{i=1,2}||\mathbf{r}_i-\mathbf{r}'_i||^2 + ||\mathrm{R}_i-\mathrm{R}'_i||^2\text,
\label{metric}
\end{align}
where the matrix norm is the Frobenius norm.
The projection algorithm for a general convex particle 
is most easily expressed in terms of the particle's support function 
$h(\mathbf{u})=\max_{\mathbf{x}\in K} \mathbf{u}\cdot\mathbf{x}$.
If the support function $h(\mathbf{u})$ of $K$ is known,
then the support function of $K_i$ is
$h_i (\mathbf{u}) = \max_{\mathbf{x}\in K} \mathbf{u}\cdot\mathrm{R}_i\mathbf{x}+\mathbf{u}\cdot\mathbf{r}_i=
 h(\mathrm{R}^T_i\mathbf{u}) + \mathbf{u}\cdot\mathbf{r}_i$
\cite{convex}.

By the separating plane theorem, $K_1$ and $K_2$ do not
overlap if and only
if a vector $\mathbf{u}$ exists such that 
$\Delta h(\mathbf{u}) = h_1 (\mathbf{u})+h_2(-\mathbf{u})\le0$
\cite{convex}. We can determine if such a vector exists by numerically minimizing
$\Delta h(\mathbf{u})/||\mathbf{u}||$,
which is bounded and attains a minimum over $\mathbf{u}$.
If the minimum value of $\Delta h(\mathbf{u})/||\mathbf{u}||$
is positive, we must make the minimal change possible to
the configuration parameters so that 
\begin{align}
h'_1 (\mathbf{u})+h'_2(-\mathbf{u})=0 \text{ for some } \mathbf{u}\text.\label{constr}
\end{align}

If we relax the condition that
$\mathrm{R}'_i$ is a rotation matrix, then we can reduce this constrained optimization
problem to a simple unconstrained optimization problem in three vector variables.
Namely, these vectors are $\mathbf{u}$,  $\mathbf{v}_1=  \mathrm{R}'^T_1 \mathbf{u}$,
and $\mathbf{v}_2=  \mathrm{R}'^T_2 \mathbf{u}$.
Given these three vectors, the new configuration parameters 
which minimize \eqref{metric} and satisfy \eqref{constr} are given by
{\allowdisplaybreaks \begin{subequations}
\begin{align}
\mathbf{r}'_1 &= \mathbf{r}_1 - \frac{\mathbf{u}}{2||\mathbf{u}||^2} 
 \Delta h \\
\mathbf{r}'_2 &= \mathbf{r}_2 + \frac{\mathbf{u}}{2||\mathbf{u}||^2} 
\Delta h \\
\mathrm{R}'_i&=\mathrm{R}_i+\frac{\mathbf{u}}{||\mathbf{u}||^2}
(\mathbf{v}_i^T-\mathbf{u}^T\mathrm{R}_i)\text,
\end{align}
where $\Delta h=h(\mathbf{v}_1) + h(-\mathbf{v}_2) + (\mathbf{r}_1-\mathbf{r}_2)\cdot\mathbf{u}$
and the configuration distance is given by
\end{subequations}}
\begin{align}
d^2=&\frac{\Delta h^2 /2 + ||\mathbf{u} \mathrm{R}_1 - \mathbf{v}_1||^2+||\mathbf{u} \mathrm{R}_2 - \mathbf{v}_2||^2}{||\mathbf{u}||^2}\text.\label{dist2}
\end{align}
And so, we have reduced the problem, as promised, to an unconstrained
minimization of \eqref{dist2} over three vector variables. Note that
\eqref{dist2} is invariant under uniform positive rescaling of 
$\mathbf{u}$,  $\mathbf{v}_1$, and $\mathbf{v}_2$,
and the resulting vanishing gradient direction must be
taken into account when performing the minimization.

The restriction on $\mathrm{R}_i$ to be a rotation matrix, 
i.e. the requirement on the rigidity of the particle, as in Ref. \cite{PDC},
is restored in the  concurrence constraint of the $D-C$ scheme.
The projection of a general matrix into the subset of orthogonal
matrices is as simple as taking its singular value decomposition
and setting all the singular values to unity \cite{projmat}.

As a method for exploring dense configurations of general hard particles,
we believe our projection-based method to be more direct and
efficient when compared
to event-driven MD simulations and stochastic MC methods
\cite{superball}. Partly,
the $D-C$ scheme draws its power from temporarily allowing
non-physical configurations, with overlaps, non-concurring replicas,
and non-rigid particles,
but then systematically acting to minimize these non-physicalities.
In MC simulations, such temporary allowance has also been observed
to be critical in efficiently exploring structures
at high density \cite{Glotzer}.

\begin{figure}
\begin{center}
\includegraphics[scale=0.75]{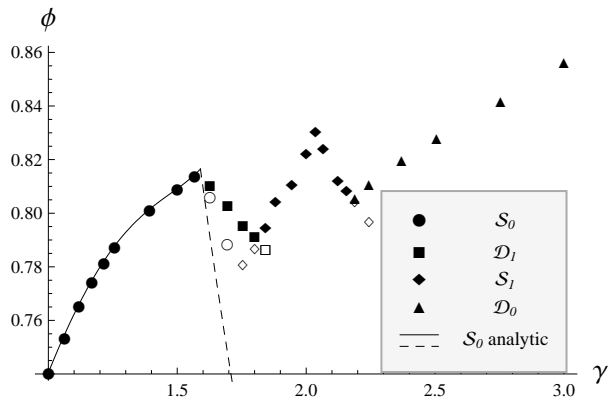}
\end{center}

\caption{\label{phis}
Highest densities $\phi$ (filled markers) achieved for
packings of tetrahedral puffs of varying
asphericity $\gamma$. The empty markers
represent the density attained by a structure at an asphericity
where it is suboptimal, as determined either by runs where the search got
trapped away from the optimal structure or by runs where the number
of particles per unit cell was incompatible with the optimal structure.
The two abrupt structural transitions can be easily seen here where
the density line for the $\mathcal{S}_1$-structure crosses the lines
for the other structures at critical asphericities at which the
$\mathcal{S}_1$-structure coexists with them.
On the other hand, the continuous structural transition between
the $\mathcal{S}_0$- and $\mathcal{D}_1$-structures is associated
with a broken symmetry instead: the line corresponds to
an analytic construction of the $\mathcal{S}_0$-structure, imposing
its tetragonal symmetry. At $\gamma\approx1.63$, next-nearest layers
come into contact with each other, leading to a sharp decline in density
(dashed line). If we allow the tetragonal symmetry to be broken,
but still allow only two particle orientations, the $D-C$ search produces 
slightly higher packing densities (empty disks). The 
$\mathcal{D}_1$-structure obtains higher densities by continuously breaking
that symmetry as well.}
\end{figure}

For a selection of puffs of different asphericities, we use the
$D-C$ scheme to perform
repeated {\it de novo} searches for periodic packings with
$p=1,2,3,4,6$, and $8$ puffs per unit cell. The densities of
the densest packings found for
each value of $\gamma$ are given in Figure \ref{phis}.
Every packing density that is reported here as putatively optimal
for a puff of some asphericity has been reproduced
by the numerical search at least 3 times from random initial
conditions and if the structure has two puffs in the primitive
unit cell, it has been reproduced in searches with both $p=2$ and $p=4$.


\begin{figure*}
\begin{center}
\includegraphics[trim=50 0 50 0,clip,scale=0.150]{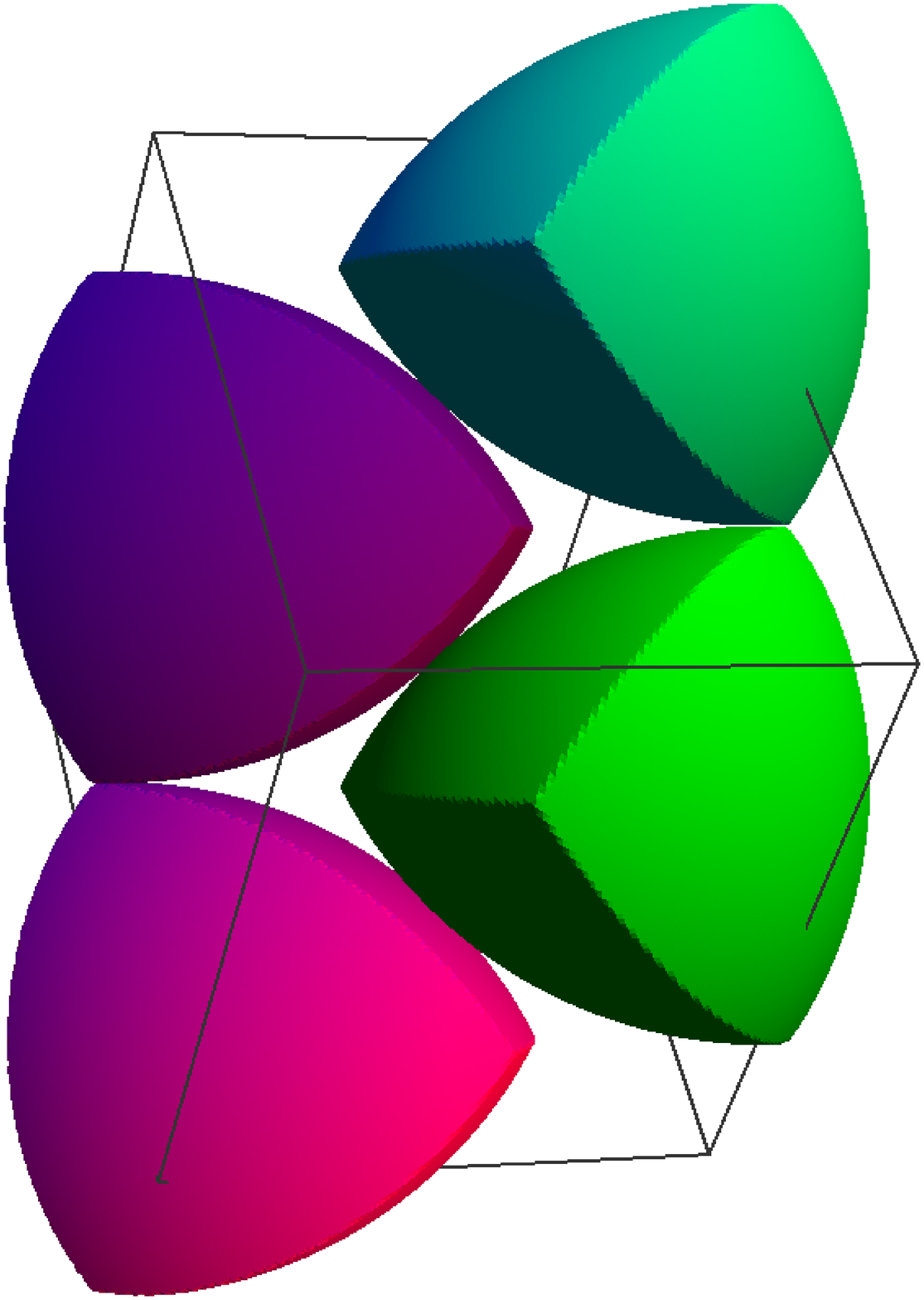}
\includegraphics[trim=50 0 50 0,clip,scale=0.150]{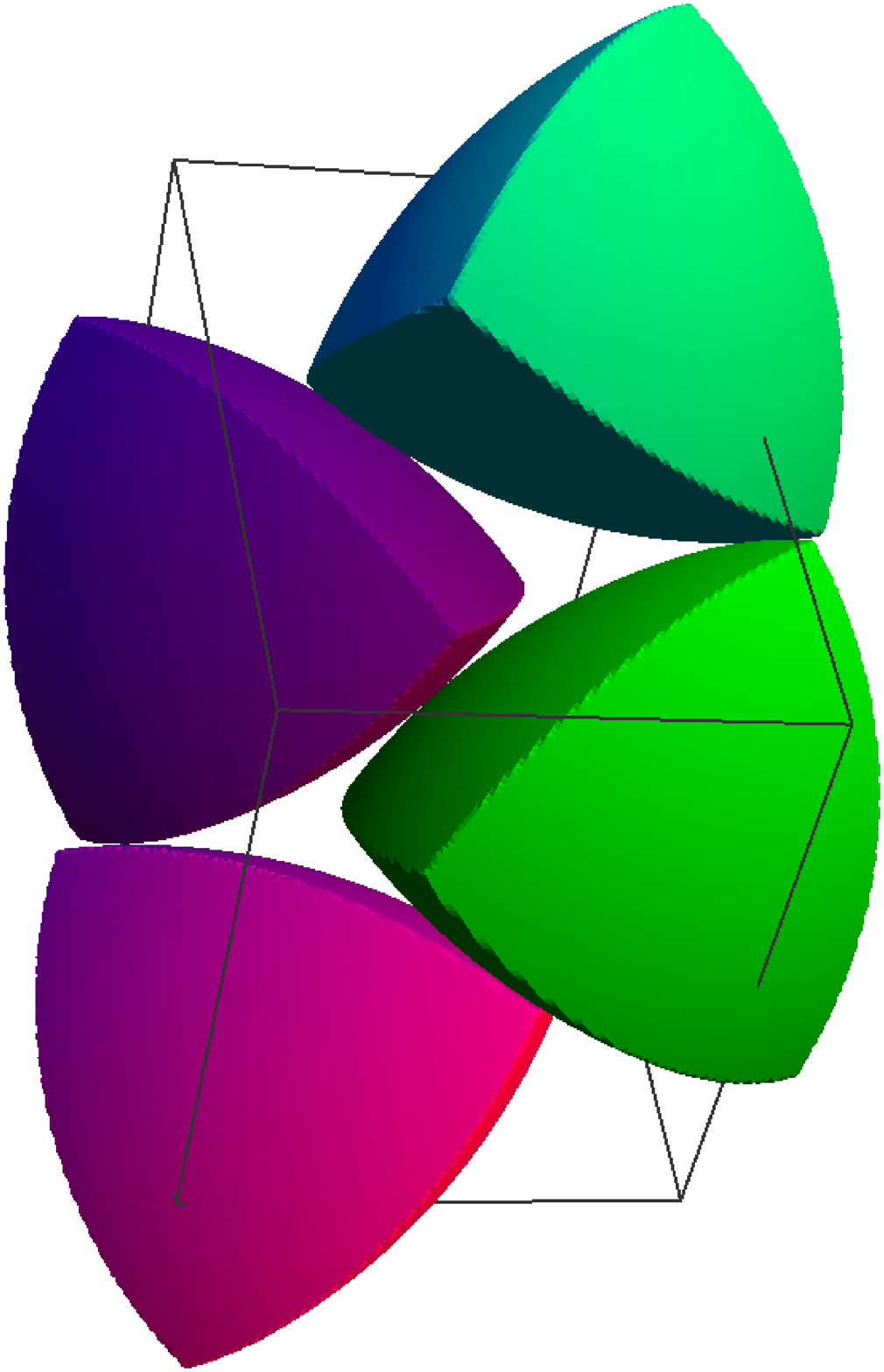}
\includegraphics[trim=110 0 50 0,clip,scale=0.190]{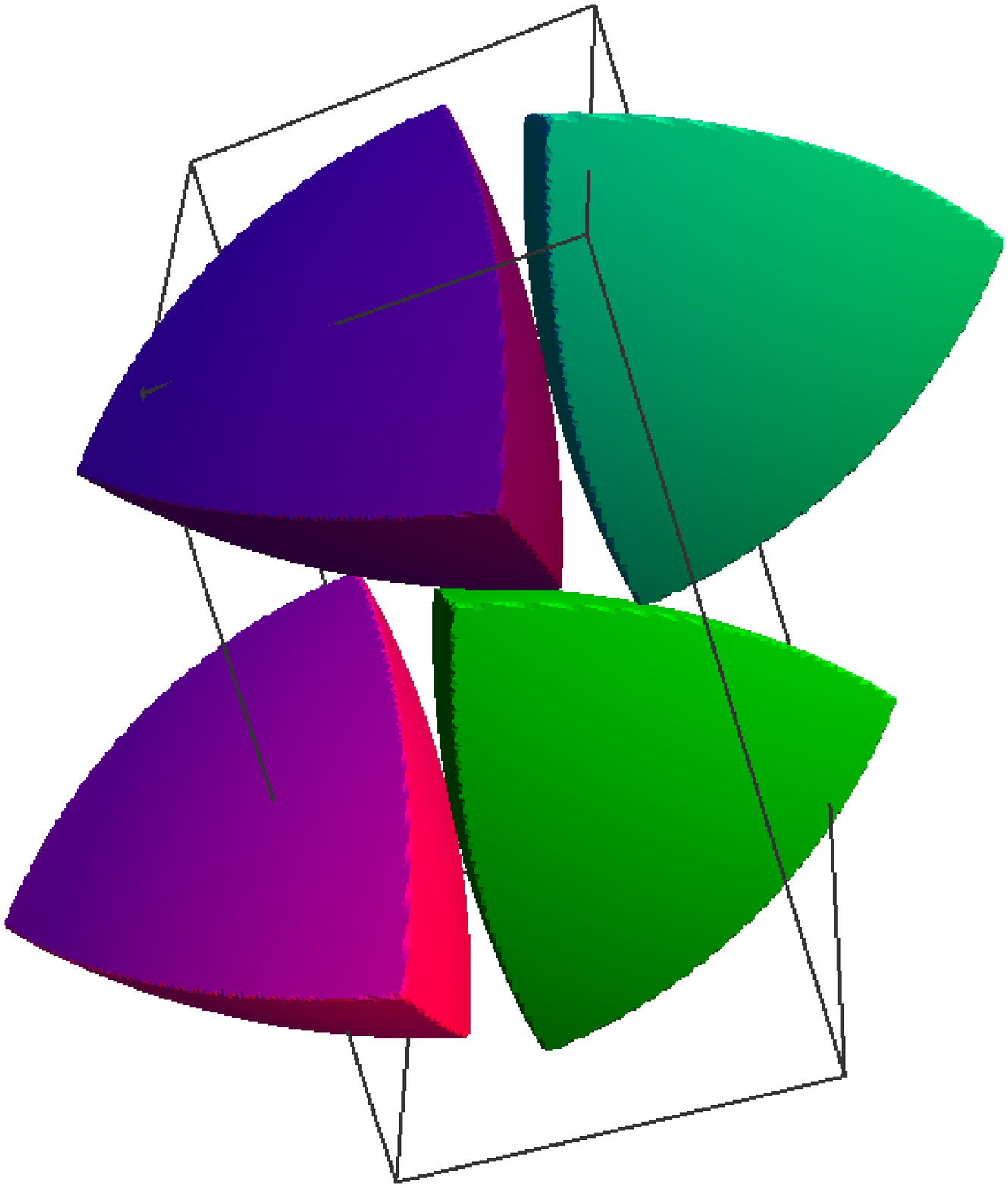}
\includegraphics[trim=140 100 50 0,clip,scale=0.250]{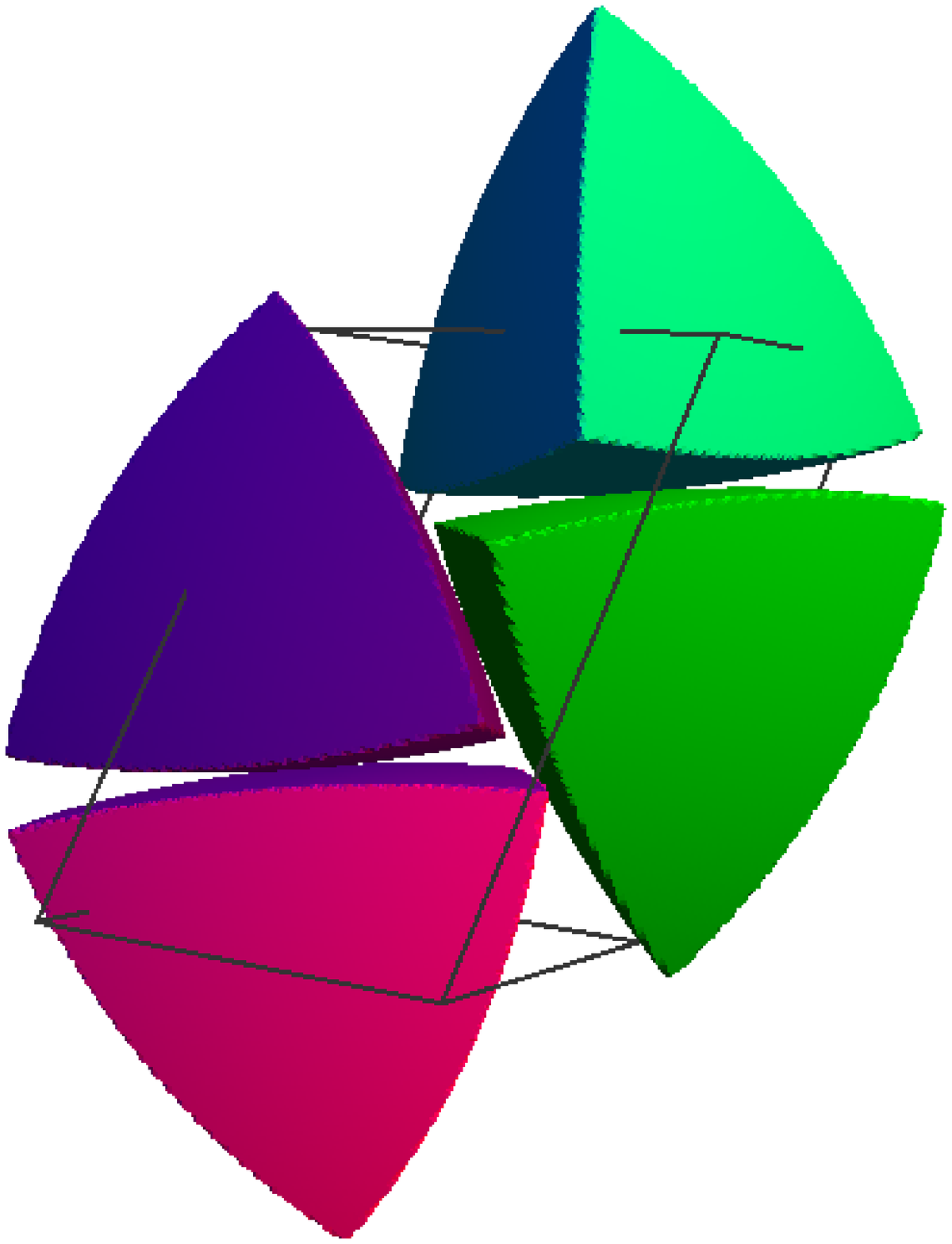}
\end{center}

\caption{\label{structs} (Color online)
a unit cell of each of the four structures described, (from left to right)
the $\mathcal{S}_0$-, $\mathcal{D}_1$-, $\mathcal{S}_1$-, and
$\mathcal{D}_0$-structures. For the $\mathcal{S}_0$- and
$\mathcal{S}_1$-structure, a unit cell consisting of two primitive
unit cells is shown. In all cases, the purple and pink puffs are
related by inversion to the green and teal puffs respectively.
The $\mathcal S_0$-structure is a body-centered tetragonal
crystal where the body-centered puff is inverted in orientation
from the corner puffs. The $\mathcal D_1$-structure occurs when
next-nearest square layers of the $\mathcal{S}_0$-structure
come into contact, and its symmetry is broken by a
re-orientation of different puffs of the same layer in different directions.
By contrast, the $\mathcal S_1$-structure
arises by an abrupt transition at both ends of the parameter interval
on which it is optimal. In the $\mathcal D_0$-structure 
the mirror planes of the green and teal puffs, which form a dimer, are
not aligned, so that both can be in contact with the purple puff.
In the limit $\gamma\to 3$, the green and teal puffs become aligned
and form a bipyramidal dimer. The structure becomes the dimer
double lattice structure reported in Refs. \cite{kallusdcg,chendcg} as
the densest known packing of regular tetrahedra.
}
\end{figure*}


As suggested by the results of our numerical searches, four different 
packing structures are optimal at different asphericities, separated
by one continuous structural transition and two abrupt ones.
Of most interest are the structures
of the optimal packing for puffs of small asphericity and large asphericity
in the parameter ranges near the sphere and the tetrahedron
respectively. The optimal packing structure for small asphericity,
which we call the $\mathcal S_0$-structure
(for {\it simple double lattice}), is a tetragonal double lattice --
that is, the union of two lattices (with one particle per unit cell)
that are related to each other by an inversion about a point
(Figure \ref{structs}) \cite{Kuperberg}. 
In the $\mathcal S_0$-structure, the puffs are arranged into square layers
so that from each layer, the puffs stick out on one side in parallel ridges
running in one direction
and on the other side in parallel ridges running in a perpendicular direction. By stacking each consecutive layer with a $90^\circ$
rotation, the ridges of one layer align with the ridges of the layer
above it. In the limit $\gamma\to 1$,
each layer approaches a square packing of spheres, and
the $\mathcal S_0$-structure approaches the f.c.c. sphere packing structure.


For large asphericities, the optimal packing structure,
which we call the $\mathcal D_0$-structure,
is a {\it dimer double lattice} (i.e. each of the two inversion-related
lattices is a lattice of dimers). The unit cell contains four puffs,
two of which, which are in contact and form a shape similar to
a triangular bipyramid, are related by inversion to the other two
(Figure \ref{structs}). We call each of the two inversion related pairs
a dimer, in analogy with the dimer double lattice of Refs. \cite{kallusdcg,
chendcg}, which is the limit of the $\mathcal D_0$-structure as $\gamma\to3$,
and is the densest known packing of regular tetrahedra. In this limit,
each dimer exactly forms a triangular bipyramid. Note that away from
$\gamma=3$, the mirror planes of the two puffs constituting the dimer
are not aligned with each other, making the dimer look twisted.
This allows both puffs to form a contact with a nearby puff (see
Figure \ref{structs}), which in the limit of the bipyramidal dimer is
facilitated simply by a contact along a common edge or vertex.


The $\mathcal S_0$-structure appears to be the optimal packing structure
from $\gamma=1$ to $\gamma\approx1.63$. On the other
end of the asphericity scale, the $\mathcal D_0$-structure
appears to be optimal from $\gamma\approx2.19$ to
$\gamma=3$. However, in the intermediate range it appears
that both of these structures are suboptimal and different
structures take over. For $\gamma\lesssim1.63$,
the density of the $\mathcal S_0$-structure increases monotonically
with asphericity. However, for the puff with $\gamma\approx1.63$,
contacts between next-nearest layers of the $\mathcal S_0$-structure appear,
and start to constrain the layer spacing. Assuming no change to the orientations
of the puffs and to the construction of the layers, this constraint leads
to a sharp drop in the density of the $\mathcal S_0$-structure.
However, the structure found by the numerical searches shows a re-orientation
of the puffs so that each layer is now composed of puffs of
two different orientations (Figure \ref{structs}). This
structure, dubbed the $\mathcal{D}_1$-structure,
still leads to a decline in the packing density, but a less dramatic one.
Therefore, we have a local maximum in the packing density at the
transition from the $\mathcal S_0$-structure to the $\mathcal D_1$-structure.
This transition seems to arise by a continuous deformation of the 
$\mathcal S_0$-structure and is not abrupt.
The re-orientation of the puffs suggests
the beginning of a tendency towards the dimerization seen in the
$\mathcal D_0$-structure. However, an intermediate structure
is encountered between the $\mathcal D_1$-structure and
the $\mathcal D_0$-structure, producing another local maximum in
the optimal density. This structure,
to be called the $\mathcal S_1$-structure, is a simple (non-dimer) double
lattice without the tetragonal symmetries of the $\mathcal S_0$-structure
(Figure \ref{structs}), and is reminiscent of the simple double lattice structure
reported in Ref. \cite{kallusdcg}. This structure appears to be separated from the
others by an abrupt transition. Figure \ref{phis} plots out the densities
of the different structures as obtained by the numerical searches and by
analytic constructions.


The proper context for the results obtained here for tetrahedral puffs is
in comparison to two other one-parameter families of particles that include
the sphere as a special case and whose dense packing structures have been
investigated vigorously, namely spheroids \cite{spheroid}
and superballs \cite{superball}. The putative
optimal packing of spheroids and superballs
becomes monotonically denser the less sphere-like they become.
By contrast, the optimal packing density
of puffs does not exhibit such monotonicity, although it is always
higher than that of the sphere (consistently with a conjecture by Ulam
that the sphere is the worst-packing three-dimensional
convex solid \cite{Ulam}). Unlike superballs, but like spheroids, the
optimal packing of puffs is in all cases (besides the sphere) not a lattice
packing, and the crystal unit cell includes at least two particles of different
orientations. However, like superballs, and unlike spheroids, the
optimal packing structure of puffs goes through an abrupt transition,
where two dissimilar structures obtain an equal, optimal density.

A major difference of tetrahedral puffs in comparison to
spheroids and superballs is the lack of inversion symmetry.
However,
not only do the putative optimal packings of puffs in all cases
have such a symmetry (Figure \ref{structs}), it is in some cases
the only symmetry of the packing besides its lattice translations. Presumably,
inversion symmetry plays an important role in forming close-packed structures
of particles with tetrahedral symmetry and maybe even of other particles,
a result already observed in the plane by Kuperberg and Kuperberg \cite{Kuperberg}.

The tetrahedral puffs exhibit a much richer optimal packing behavior than 
either spheroids or superballs, and this richness is likely to be mirrored
in the behavior of tetrahedral nano-particles. The 
variety of qualitatively different dense packing structures
observed for mathematical tetrahedra is compounded when
a physical shape parameter is added. A possible way to
experimentally access the parameter investigated here, which
describes a swollen tetrahedron, is by using colloidal particles
that swell as a function of their temperature \cite{swell}. Thus, a
variety of structures and structural transitions could be explored.
We have attempted here to provide an atlas of the possible
crystal structures which might be observed in systems of particles with
tetrahedral symmetry. 
Our candidates for optimal tetrahedral puff packings provide a starting
point for the study of the phase behavior of systems of
particles with tetrahedral symmetry, both hard and soft, away from the
limit of the mathematical tetrahedron. The method presented here, applicable
to any hard convex particle, could also be useful in characterizing possible
structures of many other particulate systems.

We acknowledge many helpful suggestions by Simon Gravel and support
from NSF grant NSF-DMR-0426568.

\bibliography{suptet}

\end{document}